\documentclass[runningheads]{llncs}
\usepackage[T1]{fontenc}
\usepackage{amsmath,amssymb,amsfonts}
\usepackage{algorithmic}
\usepackage{supertabular}
\usepackage{graphicx}
\usepackage{textcomp}
\usepackage{xcolor}
\usepackage{multirow} 
\usepackage{longtable, booktabs} 
\usepackage{longtable}
\usepackage{booktabs, makecell, tabularx, array}
\newcolumntype{L}{>{\raggedright\arraybackslash}X} 
\usepackage{stfloats} 
\usepackage{caption}  
\usepackage[T1]{fontenc}
\usepackage[utf8]{inputenc}
\usepackage{cite}
\usepackage{hyperref,graphicx,color,float}
\usepackage{float}
\usepackage{booktabs}
\usepackage{changepage} 
\def\BibTeX{{\rm B\kern-.05em{\sc i\kern-.025em b}\kern-.08em
    T\kern-.1667em\lower.7ex\hbox{E}\kern-.125emX}}
\usepackage{graphicx}
\begin{document}
\title{Code Smell Detection via Pearson Correlation and ML Hyperparameter Optimization}
\titlerunning{Code Smell Detection}
%
\author{Moinuddin Muhammad Imtiaz Bhuiyan\inst{1} \and
Kazi Ekramul Hoque\inst{1} \and
Rakibul Islam\inst{1} \and
Md. Mahbubur Rahman Tusher \inst{2} \and
Najmul Hassan \inst{3} \and
Yoichi Tomioka \inst{3} \and
Satoshi Nishimura \inst{3} \and
Jungpil Shin \inst{3} \and 
Abu Saleh Musa Miah \inst{3}
}

\institute{Dept. of Computer Science and Engineering, East Delta University, Chattogram, Bangladesh \and
Dept. of Computer Science and Engineering, Bangladesh Army University of Science and Technology, Saidpur, Bangladesh \and
Dept. of Computer Science and Engineering, The University of Aizu, Aizuwakamatsu, Japan
\email{mm.imtiaz.bhuiyan@gmail.com}\\
\email{\{musa, d8252102, ytomioka, nisim, jpshin\}@u-aizu.ac.jp}} 

\maketitle             
\begin{abstract}
This study addresses the challenge of detecting code smells in large-scale software systems using machine learning (ML). Traditional detection methods often suffer from low accuracy and poor generalization across different datasets. To overcome these issues, we propose a machine learning-based model that automatically and accurately identifies code smells, offering a scalable solution for software quality analysis. The novelty of our approach lies in the use of eight diverse ML algorithms, including XGBoost, AdaBoost, and other classifiers, alongside key techniques such as the Synthetic Minority Over-sampling Technique (SMOTE) for class imbalance and Pearson correlation for efficient feature selection. These methods collectively improve model accuracy and generalization. Our methodology involves several steps: first, we preprocess the data and apply SMOTE to balance the dataset; next, Pearson correlation is used for feature selection to reduce redundancy; followed by training eight ML algorithms and tuning hyperparameters through Grid Search, Random Search, and Bayesian Optimization. Finally, we evaluate the models using accuracy, F-measure, and confusion matrices. The results show that AdaBoost, Random Forest, and XGBoost perform best, achieving accuracies of 100\%, 99\%, and 99\%, respectively. This study provides a robust framework for detecting code smells, enhancing software quality assurance and demonstrating the effectiveness of a comprehensive, optimized ML approach.
\end{abstract}
\section{Introduction}
Strict project timelines often lead to suboptimal decisions, accumulating technical debt that necessitates code refactoring. Refactoring improves software quality and maintainability, offering long-term benefits such as better design, readability, and fault detection, though it can be resource-intensive. However, it may also introduce new flaws or alter system behavior \cite{kim2012field}. Fowler et al. \cite{fowler2018refactoring} introduced "code smells" in 1999 as indicators of poor design that lower software quality. This study treats code smell detection as a binary classification problem, aiming to identify two class-level and four method-level smells using machine learning techniques like XGB, AdaBoost, Naïve Bayes, KNN, Random Forest, Gradient Boosting, Decision Tree, and SVC \cite{miah2022movie_miah_SVM,miah2022natural_EEG_SVM_KNN}. By identifying code smells early, these models help reduce technical debt and support proactive maintenance \cite{dhanda2022machine, chaturvedi2012determining}. Despite the growing use of machine learning for code smell detection, existing methods face significant challenges, such as inconsistent classification accuracy, poor generalization to new codebases, and gaps in preprocessing and optimization. This study aims to address these issues by developing a robust and generalized framework. The novelty of this research lies not in introducing new machine learning algorithms, but in designing a comprehensive methodology that integrates and optimizes existing classifiers for code smell detection. Unlike previous studies that typically evaluate a single algorithm or conduct limited comparisons, our approach systematically assesses eight classifiers. It incorporates SMOTE for class imbalance, applies Pearson correlation-based feature selection tailored to each code smell, and uses systematic hyperparameter optimization through Grid Search, Random Search, and Bayesian Optimization, avoiding reliance on default parameters. The primary contribution is the design and empirical validation of a holistic pipeline that improves the performance of standard algorithms. This study's central objective is to compare the effectiveness of different machine learning algorithms in detecting code smells, developing a predictive framework that combines preprocessing, feature selection, and classification. The key contributions are as follows:
\begin{enumerate}
\item \textbf{Machine Learning Framework for Code Smell Detection:} We propose a machine learning-based framework to improve accuracy and generalization for code smell detection across diverse datasets. The model enhances detection in large-scale software projects by evaluating and comparing multiple algorithms.
\item \textbf{SMOTE for Data Balancing:} To address class imbalance, we apply the Synthetic Minority Over-sampling Technique (SMOTE), ensuring a balanced distribution of "smelly" and "not smelly" instances. This reduces bias and enhances the model's generalization.
\item \textbf{Feature Selection via Pearson Correlation:} We use Pearson correlation to select the most relevant features, reducing computational overhead and improving classification accuracy. For example, in the God Class dataset, key features included LOC type and WMC type, while in the Data Class dataset, features like LCOM5 and NOPK project were retained.
\item \textbf{Hyperparameter Tuning:} Hyperparameter optimization methods, including Grid Search, Random Search, and Bayesian Optimization, were used to fine-tune parameters, achieving high accuracy in AdaBoost, KNN, and Random Forest (100\%, 99\%, and 99\%, respectively).
\end{enumerate}
This research is structured into five chapters: Chapter 1 outlines the research problem and motivation; Chapter 2 provides a review of related literature; Chapter 3 details the methodology; Chapter 4 analyzes and discusses the model’s performance; and Chapter 5 concludes the work with recommendations for future research.
\section{Literature Review}
The identification of code smells in source code has become a crucial area in software refactoring. Techniques for detection range from metrics-based methods to machine learning (ML) approaches. Metrics-based methods, like those by Charalampidou et al. \cite{charalampidou2015size}, use size and cohesion metrics to detect long method smells. Rule-based systems, such as Moha et al.'s DECOR framework \cite{moha2009decor}, define explicit rules for detecting code and design smells. While ML methods are gaining attention, their application to code smell detection is still evolving. Alkharabsheh et al. \cite{alkharabsheh2019software} and Azeem et al. \cite{azeem2019machine} highlight the potential of ML to improve accuracy and efficiency in detection.
Early ML studies focused on single classifiers, such as Liu et al. \cite{liu2018deep}, who used deep learning for feature envy detection, and Barbez et al. \cite{barbez2020machine}, who explored ensemble methods for anti-patterns. Neural networks have also been used for detecting bad code smells \cite{kim2017finding, white2016deep}. Bayesian networks, as applied by Khomh et al. \cite{khomh2009bayesian} and Wang et al. \cite{wang2012can}, capture complex software patterns. Support Vector Machines (SVMs) have shown strong performance for identifying code smells \cite{maiga2012smurf, kaur2017support}. K-Nearest Neighbors (KNN), though effective, faces challenges in selecting optimal parameters; methods like GridSearchCV and robust scaling address these issues \cite{nasser2022robust}.
Innovative approaches, like the immune-inspired IDS system \cite{hassaine2010ids} and Binary Logistic Regression \cite{bryton2010reducing}, offer alternative perspectives. Studies have compared classifiers like Random Forest (RF), AdaBoost, Naïve Bayes, and KNN, using metrics such as accuracy and F-measure, with SMOTE resampling to handle data imbalance. Alsaeedi and Khan \cite{alsaeedi2019software} found RF and AdaBoost with RF to be the most effective. This study investigates six code smells, focusing on less-studied ones like Long Parameter Lists and Switch Statements, and compares eight ML classifiers—XGB, AdaBoost, Naïve Bayes, KNN, RF, Gradient Boosting, Decision Tree, and SVC—for code smell detection.
\subsection{Datasets}
This research utilizes code smell datasets compiled by Fontana et al. \cite{arcelli2016comparing}, which include labeled instances for six types of code smells: God Class, Data Class, Feature Envy, Long Method, Long Parameter List, and Switch Statements. The data were collected from 74 software systems, with each dataset comprising 420 instances categorized as either “smelly” or “not smelly” \cite{alazba2021code}.
\begin{figure}[htbp]
\begin{adjustwidth}{-2cm}{0cm}
\centering
\includegraphics[scale=0.40]{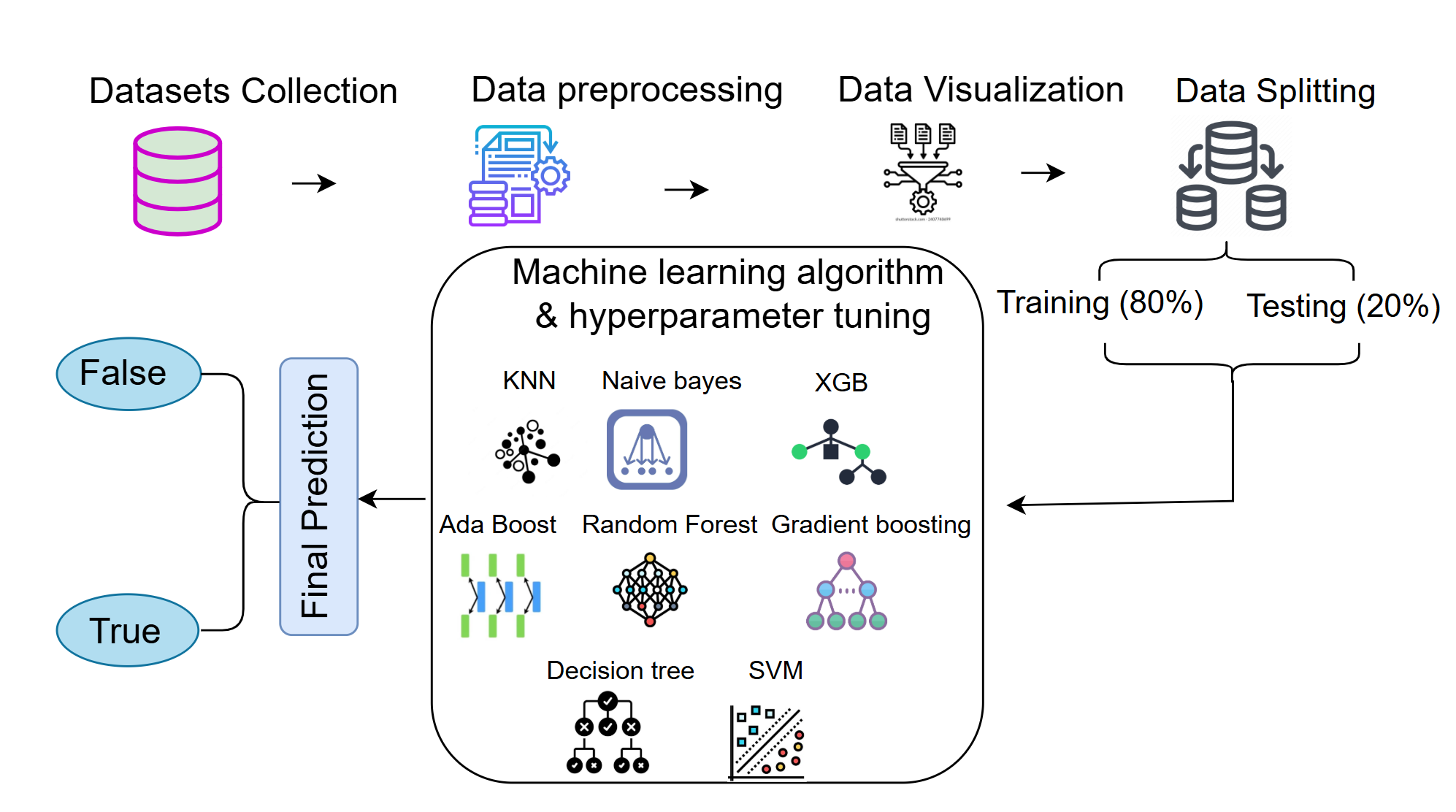}
\caption{Proposed Model Architecture}
\label{fig_main}
\end{adjustwidth}
\end{figure}
\section{Methodology}
The proposed methodology focuses on detecting code smells in software systems using machine learning (ML) models, with datasets provided by Fontana et al, which is shown in summarized in Figure \ref{fig_main}. These datasets cover six code smell types and consist of 420 labeled instances from 74 software systems. To address class imbalance, the Synthetic Minority Over-sampling Technique (SMOTE) is applied, ensuring balanced datasets. Feature selection is conducted using Pearson correlation to retain the most informative features, reducing computational complexity. Several ML algorithms, including K-Nearest Neighbors (KNN), Naïve Bayes (NB), XGBoost (XGB), AdaBoost (AB), Random Forest (RF), Gradient Boosting (GB), Decision Tree (DT), and Support Vector Machine (SVM), are evaluated. Hyperparameter optimization techniques such as Grid Search, Random Search, and Bayesian Optimization are employed to fine-tune model parameters, and performance is assessed using accuracy, precision, recall, and F1-score. This methodology provides a structured approach for effective and efficient code smell detection.

\begin{table*}[htbp!]
\centering
\caption{Best-performing hyperparameters for each dataset and model}
\label{tab:hyperparameters}
\begin{tabularx}{\textwidth}{l l L}
\toprule
\textbf{Dataset} & \textbf{Model} & \textbf{Best Hyperparameters} \\
\midrule

God Class & 
XGB & \texttt{colsample\_bytree=0.3, learning\_rate=0.1, max\_depth=3, n\_estimators=100} \\
& AB & \texttt{algorithm=SAMME, learning\_rate=1.0, n\_estimators=50} \\
& RF & \texttt{bootstrap=False, max\_depth=5, min\_samples\_leaf=2, min\_samples\_split=10, n\_estimators=100} \\
& GB & \texttt{learning\_rate=0.01, max\_depth=3, n\_estimators=200} \\
\midrule

Data Class & 
XGB & \texttt{colsample\_bytree=0.7, learning\_rate=0.1, max\_depth=3, n\_estimators=200} \\
& AB & \texttt{algorithm=SAMME, learning\_rate=1.0, n\_estimators=200} \\
& RF & \texttt{bootstrap=True, max\_depth=None, min\_samples\_leaf=1, min\_samples\_split=5, n\_estimators=50} \\
& GB & \texttt{learning\_rate=0.5, max\_depth=3, n\_estimators=100} \\
& DT & \texttt{max\_depth=5, max\_features=None, min\_samples\_leaf=1, min\_samples\_split=5} \\
\midrule

Feature Envy & 
AB & \texttt{algorithm=SAMME, learning\_rate=0.1, n\_estimators=200} \\
& RF & \texttt{bootstrap=True, max\_depth=None, min\_samples\_leaf=1, min\_samples\_split=2, n\_estimators=100} \\
& GB & \texttt{learning\_rate=1.0, max\_depth=3, n\_estimators=50} \\
\midrule

Long Method & 
KNN & \texttt{n\_neighbors=7, p=1, weights=uniform} \\
& NB & \texttt{var\_smoothing=1e-9} \\
& XGB & \texttt{colsample\_bytree=0.7, learning\_rate=0.1, max\_depth=3, n\_estimators=100} \\
& AB & \texttt{algorithm=SAMME, learning\_rate=0.1, n\_estimators=200} \\
& RF & \texttt{bootstrap=True, max\_depth=2, min\_samples\_leaf=1, min\_samples\_split=2, n\_estimators=50} \\
& GB & \texttt{learning\_rate=0.01, max\_depth=3, n\_estimators=50} \\
& DT & \texttt{max\_depth=None, max\_features=log2, min\_samples\_leaf=1, min\_samples\_split=2} \\
& SVM & \texttt{C=10, gamma=scale, kernel=linear} \\
\midrule

Long Parameter List & 
XGB & \texttt{colsample\_bytree=0.7, learning\_rate=0.3, max\_depth=3, n\_estimators=100} \\
& RF & \texttt{bootstrap=False, max\_depth=10, min\_samples\_leaf=1, min\_samples\_split=2, n\_estimators=200} \\
& GB & \texttt{learning\_rate=1.0, max\_depth=3, n\_estimators=200} \\

\addlinespace
\multicolumn{3}{>{\footnotesize\itshape}r}{Continue on the next page} \\
\bottomrule
\end{tabularx}
\end{table*}

\begin{table*}[htbp!]
\ContinuedFloat
\centering
\caption{Best-performing hyperparameters for each dataset and model (cont.)}
\begin{tabularx}{\textwidth}{l l L}
\toprule
\textbf{Dataset} & \textbf{Model} & \textbf{Best Hyperparameters} \\
\midrule

Switch Statement & 
XGB & \texttt{colsample\_bytree=0.7, learning\_rate=0.3, max\_depth=5, n\_estimators=200} \\
& RF & \texttt{bootstrap=False, max\_depth=None, min\_samples\_leaf=2, min\_samples\_split=5, n\_estimators=50} \\
& GB & \texttt{learning\_rate=1.0, max\_depth=5, n\_estimators=50} \\
& DT & \texttt{max\_depth=None, max\_features=None, min\_samples\_leaf=1, min\_samples\_split=2} \\

\bottomrule
\end{tabularx}
\end{table*}

\subsection{Preprocessing}
The datasets exhibit class imbalance, with unequal representation of “smelly” and “not smelly” instances. For example, the God Class dataset contains 140 “smelly” and 280 “not smelly” instances, and similar imbalances exist in the other datasets: Data Class (140/280), Feature Envy (140/280), Long Method (140/280), Long Parameter List (138/282), and Switch Statements (129/291).  To address this issue and enhance model performance, the Synthetic Minority Over-sampling Technique (SMOTE) was applied. SMOTE generates synthetic samples for the minority class, resulting in balanced datasets with 280/280 instances for most code smells, and 282/282 for Long Parameter List and 291/291 for Switch Statements. This preprocessing step improved the representativeness of the datasets and the predictive accuracy of the machine learning models.

\subsection{Balance with Synthetic Minority Over-sampling Technique (SMOTE)}
To address class imbalance in the datasets, the Synthetic Minority Over-sampling Technique (SMOTE) was employed. Unlike simple duplication, SMOTE is a sophisticated oversampling method that generates synthetic minority class instances by interpolating between existing ones in the feature space. This process effectively broadens the decision region for the minority class, enabling classifiers to learn more robust boundaries and significantly reducing model bias toward the majority class, which is crucial for accurate code smell detection.  

\subsection{Feature Selection Using Pearson Correlation}
Feature selection was performed using Pearson correlation to reduce redundancy and improve model efficiency. By calculating the correlation between each feature and the target variable, only features with correlations above the dataset's average were retained. This method is particularly valuable as it ensures the most relevant features are selected, optimizing the predictive performance of the model. The novelty of this approach lies in its tailored application to each code smell type, where unique features were selected based on their specific correlation values. For example, in the God Class dataset, features like LOC type and WMC type were selected, while for Data Class, features like LCOM5 and NOPK project were prioritized. Similar tailored feature selection was applied to the other code smells: Feature Envy (FDP method, ATFD method), Long Method (MAXNESTING method, LOC method), Long Parameter List (NOP method, LOC method), and Switch Statements (LOC method, MAXNESTING method). These selected features formed the foundation for training and evaluating the machine learning models, ensuring efficient and accurate detection of code smells in the later stages.

\subsection{Machine Learning Algorithms}  
This study employs eight supervised machine learning algorithms to enable a comprehensive comparative analysis. The selected models include fundamental algorithms like K-Nearest Neighbors, Naive Bayes, and Decision Trees, alongside more complex methods such as Support Vector Machines \cite{miah2025emg,kabir2024exploring_miah_LDA,miah2025methodologica_pd,10937116_pd_matsumoto}. Advanced ensemble techniques, including Random Forest, AdaBoost, Gradient Boosting, and XGBoost, are also included. This diverse selection ensures a thorough evaluation of classification performance across different learning paradigms, covering instance-based, probabilistic, tree-structured, and maximum-margin classifiers, as well as bagging and boosting ensembles.
To develop robust and generalizable models, all classifiers were trained and validated systematically. Key hyperparameters were tuned using Grid Search, Random Search, and Bayesian Optimization to identify the optimal configuration for each algorithm-dataset pair, aiming to minimize overfitting and achieve a balanced bias-variance trade-off. After tuning, the final models were retrained on the combined training and validation sets. Performance was rigorously evaluated on a held-out test set to obtain unbiased estimates of predictive accuracy and generalization to unseen data.

\section{Result Analysis}
In this study, six distinct types of code smells were detected: God Class, Data Class, Feature Envy, Long Method, Long Parameter List, and Switch Statements. For each dataset, the data was split into training and testing sets, with 80\% allocated for training and 20\% for testing. Specifically, the God Class, Data Class, Feature Envy, and Long Method datasets each had 392 instances for training and 168 for testing. The Long Parameter List dataset had 394 instances for training and 170 for testing, while the Switch Statements dataset consisted of 407 training instances and 175 testing instances. To predict code smells, we applied several machine learning algorithms, including K-Nearest Neighbors (KNN), Naïve Bayes, XGBoost, AdaBoost, and Random Forest. The performance of these models was evaluated using key metrics such as accuracy, confusion matrix, precision, and recall to determine the best-performing model.

\subsection{Model Performance without Hyperparameter Optimization}  
Table~\ref{tab:all_results} presents the accuracy, precision, recall, and F1-score for eight machine learning models across six code smell categories: God Class, Data Class, Feature Envy, Long Method, Long Parameter List, and Switch Statements. Tree-based ensemble methods—XGBoost (XGB), Random Forest (RF), and Gradient Boosting (GB)—consistently delivered top performance. In the God Class category, all three methods achieved perfect scores (100\% across all metrics). For the Data Class category, XGB led with 98\% accuracy, while Naïve Bayes (NB) had high recall (99\%) but lower precision (67\%). In the Feature Envy category, RF and GB excelled with 96\% accuracy and F1-score. Most models performed near-perfectly for Long Method. In the Long Parameter List category, RF and GB (94\%) outperformed KNN (89\%) and SVM (90\%), while AdaBoost lagged slightly. For Switch Statements, XGB, RF, GB, and Decision Tree (DT) achieved perfect recall and F1-scores of 97\%. Overall, XGB, RF, and GB were the most consistent across all categories, while Naïve Bayes showed variability—strong recall but weaker precision. AdaBoost and Decision Tree remained competitive in most cases.
\begin{table*}[htbp]
\centering
\caption{Performance Comparison of Machine Learning Models for Different Code Smell Categories}
\label{tab:all_results}
\makebox[\textwidth][c]{%
\renewcommand{\arraystretch}{.8}
\setlength{\tabcolsep}{1pt}
\scriptsize

\begin{tabular}{|l|c c c c|c c c c|c c c c|c c c c|c c c c|c c c c|}
\hline
\multirow{2}{*}{\textbf{Model}} & \multicolumn{4}{c|}{\textbf{God Class}} & \multicolumn{4}{c|}{\textbf{Data Class}} & \multicolumn{4}{c|}{\textbf{Feature Envy}} & \multicolumn{4}{c|}{\textbf{Long Method}} & \multicolumn{4}{c|}{\textbf{Long Param. List}} & \multicolumn{4}{c|}{\textbf{Switch Statements}} \\
\cline{2-25}
 & Acc & Prec & Rec & F1 & Acc & Prec & Rec & F1 & Acc & Prec & Rec & F1 & Acc & Prec & Rec & F1 & Acc & Prec & Rec & F1 & Acc & Prec & Rec & F1 \\
\hline
KNN & 98.00 & 100.00 & 96.00 & 98.00 & 93.00 & 88.00 & 99.00 & 93.00 & 94.00 & 94.00 & 96.00 & 95.00 & 99.00 & 99.00 & 100.00 & 99.00 & 89.00 & 92.00 & 85.00 & 89.00 & 92.00 & 90.00 & 94.00 & 92.00 \\
NB & 97.00 & 100.00 & 95.00 & 97.00 & 77.00 & 67.00 & 99.00 & 80.00 & 88.00 & 92.00 & 86.00 & 89.00 & 98.00 & 95.00 & 100.00 & 98.00 & 81.00 & 89.00 & 70.00 & 78.00 & 88.00 & 90.00 & 84.00 & 87.00 \\
XGB & \textbf{100.00} & \textbf{100.00} & \textbf{100.00} & \textbf{100.00} & 98.00 & 96.00 & 99.00 & 97.00 & 95.00 & 94.00 & 98.00 & 96.00 & 99.00 & 99.00 & 100.00 & 99.00 & 94.00 & 94.00 & 93.00 & 93.00 & \textbf{97.00} & 93.00 & \textbf{100.00} & \textbf{97.00} \\
AdaBoost & 99.00 & 100.00 & 98.00 & 99.00 & 96.00 & 93.00 & 99.00 & 96.00 & \textbf{96.00} & \textbf{96.00} & 97.00 & \textbf{96.00} & 99.00 & 99.00 & 100.00 & 99.00 & 91.00 & 91.00 & 89.00 & 90.00 & 91.00 & 85.00 & 99.00 & 91.00 \\
RF & \textbf{100.00} & \textbf{100.00} & \textbf{100.00} & \textbf{100.00} & 96.00 & 94.00 & 99.00 & 96.00 & \textbf{96.00} & 94.00 & \textbf{99.00} & \textbf{96.00} & 99.00 & 99.00 & 100.00 & 99.00 & \textbf{94.00} & 95.00 & \textbf{93.00} & \textbf{94.00} & \textbf{97.00} & 93.00 & \textbf{100.00} & \textbf{97.00} \\
GB & \textbf{100.00} & \textbf{100.00} & \textbf{100.00} & \textbf{100.00} & 97.00 & 95.00 & 99.00 & 97.00 & \textbf{96.00} & 94.00 & \textbf{99.00} & \textbf{96.00} & 99.00 & 99.00 & 100.00 & 99.00 & \textbf{94.00} & 96.00 & 90.00 & 93.00 & 96.00 & 92.00 & \textbf{100.00} & 96.00 \\
DT & 98.00 & 99.00 & 98.00 & 98.00 & 96.00 & 94.00 & 99.00 & 96.00 & 95.00 & 94.00 & 98.00 & 96.00 & 99.00 & 99.00 & 100.00 & 99.00 & 93.00 & 93.00 & 93.00 & 93.00 & \textbf{97.00} & \textbf{94.00} & \textbf{100.00} & \textbf{97.00} \\
SVM & 98.00 & 100.00 & 96.00 & 98.00 & 91.00 & 84.00 & 99.00 & 91.00 & 94.00 & 94.00 & 96.00 & 95.00 & 99.00 & 99.00 & 100.00 & 99.00 & 90.00 & 90.00 & 89.00 & 90.00 & 91.00 & 89.00 & 92.00 & 90.00 \\
\hline
\end{tabular}}
\end{table*}
\subsection{Model Performance with Hyperparameter Optimization}  
To enhance model performance, we applied hyperparameter optimization across all machine learning algorithms. This significantly improved accuracy, recall, and F1-scores, particularly for challenging categories like Long Parameter List and Switch Statements (Table~\ref{tab:hyper_results}). In God Class, AdaBoost achieved perfect scores, followed closely by KNN, XGB, RF, and GB. For Data Class, GB led with 98\% accuracy, while NB maintained high recall (99\%) but low precision (67\%). In Feature Envy, SVM reached 97\% accuracy, with RF, GB, and DT at 96\%. Long Method remained the easiest category to detect, with nearly all models scoring between 99\% and 100\%. In Long Parameter List, GB led with 94\%, followed by XGB (93\%) and RF (90\%). For Switch Statements, XGB, RF, GB, and DT achieved perfect recall and F1-scores close to 96\%. Overall, ensemble methods—particularly GB, XGB, and AdaBoost—provided the most balanced and high-performing results. KNN and SVM also improved, while NB continued to excel in recall but struggled with precision.

\begin{table*}[htbp]
\centering
\caption{Result Analysis of All Code Smell Classes after Hyperparameter Optimization}
\label{tab:hyper_results}
\makebox[\textwidth][c]{%
\renewcommand{\arraystretch}{.8}
\setlength{\tabcolsep}{1pt}
\scriptsize
\begin{tabular}{|l|c c c c|c c c c|c c c c|c c c c|c c c c|c c c c|}
\hline
\multirow{2}{*}{\textbf{Model}} & \multicolumn{4}{c|}{\textbf{God Class}} & \multicolumn{4}{c|}{\textbf{Data Class}} & \multicolumn{4}{c|}{\textbf{Feature Envy}} & \multicolumn{4}{c|}{\textbf{Long Method}} & \multicolumn{4}{c|}{\textbf{Long Param. List}} & \multicolumn{4}{c|}{\textbf{Switch Statements}} \\
\cline{2-25}
& Acc & Prec & Rec & F1 & Acc & Prec & Rec & F1 & Acc & Prec & Rec & F1 & Acc & Prec & Rec & F1 & Acc & Prec & Rec & F1 & Acc & Prec & Rec & F1 \\
\hline
KNN & 99.00 & 100.00 & 99.00 & 99.00 & 97.00 & 94.00 & 100.00 & 97.00 & 95.00 & 93.00 & 99.00 & 96.00 & 99.00 & 99.00 & 100.00 & 99.00 & 92.00 & 91.00 & 94.00 & 92.00 & 95.00 & 91.00 & 100.00 & 95.00 \\
NB & 97.00 & 100.00 & 95.00 & 97.00 & 77.00 & 67.00 & 99.00 & 80.00 & 88.00 & 92.00 & 86.00 & 89.00 & 98.00 & 95.00 & 100.00 & 98.00 & 81.00 & 89.00 & 70.00 & 78.00 & 88.00 & 90.00 & 84.00 & 87.00 \\
XGB & 99.00 & 100.00 & 99.00 & 99.00 & 97.00 & 95.00 & 99.00 & 97.00 & 95.00 & 94.00 & 98.00 & 96.00 & 99.00 & 99.00 & 100.00 & 99.00 & 94.00 & 95.00 & 91.00 & 93.00 & 97.00 & 93.00 & 100.00 & 97.00 \\
AdaBoost & 100.00 & 100.00 & 100.00 & 100.00 & 97.00 & 95.00 & 99.00 & 97.00 & 95.00 & 96.00 & 96.00 & 96.00 & 99.00 & 99.00 & 100.00 & 99.00 & 89.00 & 96.00 & 82.00 & 88.00 & 94.00 & 91.00 & 98.00 & 94.00 \\
RF & 99.00 & 99.00 & 100.00 & 99.00 & 96.00 & 93.00 & 100.00 & 96.00 & 96.00 & 94.00 & 99.00 & 96.00 & 99.00 & 98.00 & 100.00 & 99.00 & 91.00 & 95.00 & 87.00 & 90.00 & 96.00 & 92.00 & 100.00 & 96.00 \\
GB & 99.00 & 99.00 & 100.00 & 99.00 & 98.00 & 96.00 & 99.00 & 97.00 & 96.00 & 95.00 & 98.00 & 96.00 & 99.00 & 99.00 & 100.00 & 99.00 & 94.00 & 95.00 & 93.00 & 94.00 & 96.00 & 92.00 & 100.00 & 96.00 \\
DT & 98.00 & 98.00 & 98.00 & 98.00 & 95.00 & 91.00 & 97.00 & 94.00 & 96.00 & 96.00 & 97.00 & 96.00 & 99.00 & 98.00 & 100.00 & 99.00 & 90.00 & 96.00 & 83.00 & 89.00 & 97.00 & 94.00 & 100.00 & 97.00 \\
SVM & 98.00 & 100.00 & 96.00 & 98.00 & 93.00 & 87.00 & 99.00 & 93.00 & 97.00 & 96.00 & 99.00 & 97.00 & 99.00 & 98.00 & 100.00 & 99.00 & 90.00 & 90.00 & 89.00 & 90.00 & 93.00 & 91.00 & 95.00 & 93.00 \\
\hline
\end{tabular}}
\end{table*}
\subsection{State of the Art Comparison}
We compare the proposed method with the state-of-the-art techniques, as shown in Table~\ref{tab:all_comparsion}. The results indicate that the proposed Gradient Boosting (GB) model outperforms the Stack-SVM model in most categories. For God Class, the proposed method achieved a perfect recall of 100\%, leading to a high F1-score of 99\%. In Data Class, Feature Envy, and Long Method, the proposed method also maintained competitive performance, with F1-scores of 97\%, 96\%, and 99\%, respectively. Notably, the Stack-SVM model achieved lower performance in several categories, with particularly weak results in Long Parameter List and Switch Statements. 

\begin{table*}[htbp]
\centering
\caption{Performance Comparison of proposed Models for Different Code Smell Categories with state-of-the-art method}
\label{tab:all_comparsion}
\makebox[\textwidth][c]{%
\renewcommand{\arraystretch}{.8}
\setlength{\tabcolsep}{1pt}
\scriptsize

\begin{tabular}{|l|c c c c|c c c c|c c c c|c c c c|c c c c|c c c c|}
\hline
\multirow{2}{*}{\textbf{Model}} & \multicolumn{4}{c|}{\textbf{God Class}} & \multicolumn{4}{c|}{\textbf{Data Class}} & \multicolumn{4}{c|}{\textbf{Feature Envy}} & \multicolumn{4}{c|}{\textbf{Long Method}} & \multicolumn{4}{c|}{\textbf{Long Param. List}} & \multicolumn{4}{c|}{\textbf{Switch Statements}} \\
\cline{2-25}
 & Acc & Prec & Rec & F1 & Acc & Prec & Rec & F1 & Acc & Prec & Rec & F1 & Acc & Prec & Rec & F1 & Acc & Prec & Rec & F1 & Acc & Prec & Rec & F1 \\
\hline

Proposed (GB) & 99.00 & 99.00 & 100.00 & 99.00 & 98.00 & 96.00 & 99.00 & 97.00 & 96.00 & 95.00 & 98.00 & 96.00 & 99.00 & 99.00 & 100.00 & 99.00 & 94.00 & 95.00 & 93.00 & 94.00 & 96.00 & 92.00 & 100.00 & 96.00  \\\hline
Stack-SVM\cite{alazba2021code} & 97.0 & - & - & 95.49 & 98.79 & - &- & 98.17 & 94.55 & - & - & 91.90 & 99.24 & - & - & 98.87 & 92.10 & - & - & 87.69 & 88.19 & -& - & 80.50 \\
\hline
\end{tabular}
}
\end{table*}


\section{Conclusion and Future Work}
This study addresses the challenge of detecting code smells in software systems by proposing a machine learning-based model that enhances detection accuracy and generalization. Traditional methods often struggle with low accuracy and weak generalization across diverse datasets, which our approach seeks to overcome. The novelty of our model lies in the use of eight diverse machine learning algorithms, including advanced ensemble techniques like XGBoost and AdaBoost, ensuring robust evaluation. We also applied the Synthetic Minority Over-sampling Technique (SMOTE) to address class imbalance and used Pearson correlation for efficient feature selection, enhancing the model's performance and reducing computational complexity. The process involved acquiring labeled datasets for six code smells, preprocessing the data with SMOTE, selecting features using Pearson correlation, and training the algorithms with hyperparameter optimization techniques. Model evaluation was performed using accuracy, F-measure, and confusion matrices, revealing strong performance, particularly from AdaBoost, Random Forest, and XGBoost. Despite its success, the model has limitations, such as challenges in parameter selection for some classifiers, like K-Nearest Neighbors. Future work we plan to focus on incorporating deep learning models, expanding the dataset, and exploring advanced techniques like transfer learning to improve predictive capabilities and generalization across diverse software systems.

\bibliography{Ref}
\end{document}